\documentclass[conference]{IEEEtran}
\IEEEoverridecommandlockouts
\usepackage{cite}
\usepackage{amsmath,amssymb,amsfonts}
\usepackage{algorithmic}
\usepackage[dvipdfmx]{graphicx}
\usepackage{textcomp}
\usepackage{here}
\usepackage{url} 
\usepackage{multicol}
\usepackage{multirow}

\newcommand{\bhline}[1]{\noalign{\hrule height #1}}  
\newcommand{\vect}[1]{{\mbox{\boldmath $#1$}}}
\usepackage[linesnumbered,ruled,vlined]{algorithm2e}

\def\BibTeX{{\rm B\kern-.05em{\sc i\kern-.025em b}\kern-.08em
    T\kern-.1667em\lower.7ex\hbox{E}\kern-.125emX}}
\begin{document}

\title{Distributed collaborative anomalous sound detection by embedding sharing}

\author{\IEEEauthorblockN{Kota Dohi}
  \IEEEauthorblockA{\textit{Hitachi Ltd.} \\
    Tokyo, Japan \\
    kota.dohi.gr@hitachi.com}
  \and
  \IEEEauthorblockN{Yohei Kawaguchi}
  \IEEEauthorblockA{\textit{Hitachi Ltd.} \\
    Tokyo, Japan \\
    yohei.kawaguchi.xk@hitachi.com}}

\maketitle

\begin{abstract}
To develop a machine sound monitoring system, a method for detecting anomalous sound is proposed. In this paper, we explore a method for multiple clients to collaboratively learn an anomalous sound detection model while keeping their raw data private from each other. In the context of industrial machine anomalous sound detection, each client possesses data from different machines or different operational states, making it challenging to learn through federated learning or split learning. In our proposed method, each client calculates embeddings using a common pre-trained model developed for sound data classification, and these calculated embeddings are aggregated on the server to perform anomalous sound detection through outlier exposure. Experiments showed that our proposed method improves the AUC of anomalous sound detection by an average of 6.8\%.
\end{abstract}


\section{Introduction}
\label{sec:intro}
To address the worldwide shortage of maintenance workers, 
researchers have been investigating anomalous sound detection (ASD) methods, which 
detect anomalies in industrial machines by sound
 \cite{Koizumi2017, Suefusa2020, mnasri2022anomalous}.

In practice, unsupervised ASD methods are frequently employed due to difficulties in collecting a large number of anomalous samples and capturing all anomaly patterns. However, these methods may result in lower detection accuracy, potentially failing to meet the level of accuracy required.

One recognized method to improve the accuracy of unsupervised ASD models is outlier exposure  \cite{Giri2020}, which leverages data from multiple machines or various operational states.
However, in many practical cases, collecting a sufficient amount of data for outlier exposure by one customer poses a significant challenge. The process of outlier exposure involves learning to distinguish between data from distinct machines or conditions. Consequently, this learning process is impractical when a limited number of machines with a limited number of operational states are available.
While data from multiple customers can be utilized, the presence of sensitive information in data makes it challenging. 

Distributed collaborative machine learning (DCML) methods \cite{zhao2018federated} such as federated learning \cite{mcmahan2016communication} and split learning \cite{vepakomma2018split} allow for the utilization of data from multiple clients while preserving data confidentiality. Yet, if the data distribution varies significantly between clients, the accuracy of the learned model could considerably decrease.

In this paper, we explore an approach of utilizing data embeddings gathered from multiple clients. Embeddings are dimensionally reduced representations of raw data produced by trained models, and are not human-interpretable. Also, embeddings preserve valuable information related to the task for which the model that calculated the embedding was trained. Therefore, depending on the relationship between the task and the one multiple clients collaboratively want to accomplish, embeddings can be used instead of raw data. We propose to share embeddings from pre-trained model at each client and utilize the shared embeddings for ASD through outlier exposure. Experimental results using the DCASE2020 Task2 development dataset with our proposed approach showed an average improvement of 6.8\% in ASD performance compared to training unsupervised models using only each client's data.

\section{Problem Statement}
\label{sec:statement}
ASD involves determining the state of a machine - normal or anomalous - based on an anomaly score derived from the machine's sound by a trained model. Each piece of sound data is categorized as anomalous if its anomaly score surpasses a threshold value. We focus on unsupervised ASD, a scenario where only normal sound data is accessible for training.

In this study, we consider a setup with a single server and multiple clients. Each client holds data from a unique machine operating under a distinct condition. This assumption is based on practical difficulties often faced in preparing multiple machines or collecting data under various operating conditions. It is important to note that we assume no two clients possess the same type of machine or identical operating conditions. The primary objective of our research is to develop a highly accurate ASD model for each client's data while preventing the leakage of raw data between clients.  

\section{Relation to prior work}
\subsection{ASD without anomalous samples}
Unsupervised ASD methods are mainly used when anomalous samples are limited. These methods typically train generative models such as autoencoder (AE), variational autoencoder (VAE) \cite{Kingma2014a}, and 
normalizing flows (NF) \cite{Tabak2013, Kingma2018} to yield higher likelihoods for normal data
and lower likelihoods for anomalous data.
However, it has beeen reported that these models may ``overlook'' anomalies and assign higher likelihoods to anomalous data \cite{Koizumi2020spider}.

If data from multiple machines or different operating conditions are available, detection performance can be enhanced through the application of outlier exposure. This process involves training a model to differentiate data from various machines or conditions, facilitating a more precise representation of the normal data distribution. 
However, implementing outlier exposure can be challenging if data from only a single machine with unique operating condition is accessible.

Our proposed method enables the application of outlier exposure even in scenarios where each client possesses a single machine under a unique operating condition. This can be achieved by aggregating the data embedding from each client on a central server.

\subsection{Distributed collaborative machine learning (DCML)}
In DCML, the goal is for distributed clients and servers to collaborate on learning without exposing each client's data.

Federated learning is a known DCML method that shares the weights of the model trained on data of each client. In federated learning, each client sends the weights of the model to a central server. The server then combines the model weights obtained from each client and sends them back. This process facilitates the training of high-accuracy models without sharing raw data between clients.

Federated learning works when the distribution of data across clients matches the distribution of data across all clients (Independent and Identically Distributed, IID). However, it is known to be challenging when the distribution of data differs significantly across clients (non-IID) \cite{zhao2018federated}. In the context of unsupervised ASD, it is common for different clients to have different machines or machines in different conditions. Hence, the scenarios where federated learning can be effectively applied are limited.

Split learning is another DCML method that involves dividing the network into a server-side network and a client-side network. The loss is computed using the server-side network on the embeddings calculated with the client-side network, and both the client-side and server-side networks are updated using the backpropagation method. This allows learning to be performed without sharing raw data among clients. However, each model update is performed with embeddings from only one client, and it has been found to be less capable of handling non-IID data compared to federated learning \cite{gao2020endtoend}.

Our proposed method is similar to split learning, but uses a pre-trained model as the client-side network and do not conduct learning on the client side. As a result, the client-side pre-trained model is not updated. This strategy allows us to use the embeddings of all clients for each model update. Consequently, even if each client has data from only one machine operating under one condition, we can still perform learning.

\section{Distributed collaborative ASD with embedding sharing}
\label{sec:conventional}
\subsection{Overview of the proposed method}

\begin{algorithm}[t]  
    \SetAlgoLined  
    \DontPrintSemicolon  
    \caption{Collaborative ASD by embedding sharing}\label{alg:fed_anomaly_detection}  
    \SetAlgoVlined  
    \SetKw{In}{in}  
    \SetKw{Continue}{continue}  
    \KwIn{Pairs of train data and corresponding IDs $(\mathbf{X}_{k}, \mathbf{S}_{k})=(\left\{\vect{x}_{kn}\right\}_{n=1}^N, \left\{s_k\right\}_{n=1}^N)$ at each client $k$,
    pairs of test data and IDs $(\mathbf{X'}_{k}, \mathbf{S'}_{k})=(\left\{\vect{x'}_{kn}\right\}_{n=1}^{N'}), \left\{s_k\right\}_{n=1}^{N'})$ at each client $k$,
    common pre-trained model $f_{c}$ at all clients,
    anomaly detection model $f_{a, k}$ at each client,
    ID-classification model $f_{s}$ at the server}  
    \KwOut{Set of anomaly scores $\mathbf{A}_k$ for each client $k$}  
    \BlankLine  
    \tcc{Training Phase}  
    \tcc{Clients calculate embeddings $\mathbf{E}_{k}$}  
    \For{each client $k$ in parallel}{ 
        \tcp{Calculate embeddings $\mathbf{E}_{k}$}
        $\mathbf{E}_{k} \leftarrow f_c (\mathbf{X}_{k})$ \\
        Send embeddings $\mathbf{E}_{k}$ and IDs $\mathbf{S}_k$ to the server\\  
    }  
    \BlankLine
    \tcc{Server trains ID-classification model $f_{s}$}  
    Server receives embeddings $\mathbf{E}_{k}$ and IDs $\mathcal{S}_k$ from each client\\  
    Server trains ID-classification model $f_{s}$ to predict IDs from embeddings $\mathbf{E}_{k}$\\  
    Server sends trained $f_{s}$ to all clients\\  
    \BlankLine
    \tcc{Clients train anomaly detection model $f_a$}
    \For{each client $c$ in parallel}{  
        Receive trained $f_{s}$ from server\\
        \tcp{Calculate embeddings $\mathbf{\Tilde{E}}_{k}$}
        $\mathbf{\Tilde{E}}_{k} \leftarrow f_s (\mathbf{E}_{k})$ \\
        Train an anomaly detection model $f_{a,k}$ using embeddings $\mathbf{\Tilde{E}}_{k}$\\  
    }  
    \tcc{Inference Phase}  
    \BlankLine  
    \For{each client $c$ in parallel}{
        $\mathbf{E}'_{k} \leftarrow f_c (\mathbf{X}'_{k})$ \\
        $\mathbf{\Tilde{E}'}_{k} \leftarrow f_s (\mathbf{E}'_{k})$ \\
        \tcp{Calculate anomaly scores using the anomaly detection model $f_a$}
        $\mathbf{A}_{k} \leftarrow f_a (\mathbf{\Tilde{E}'}_{k})$ \\
    }  
\end{algorithm}  

In Algorithm \ref{alg:fed_anomaly_detection}, we present the process for each client and server in the proposed method. 

We assume that each client has pairs of data and corresponding IDs for both training $(\mathbf{X}_k, \mathbf{S}_k)$ and testing $(\mathbf{X}'_k, \mathbf{S}'_k)$. Here, an ID is a label that specifies the particular class each sample of sound data belongs to. This particular class could correspond to an individual machine or a certain operating state of the machine. Note that, because we assume each client has only one machine in a single operational state, all samples from client $k$ correspond to the same ID $s_k$. A common pre-trained model $f_c$ is distributed to all clients. This model is designed to generate embeddings from raw data that hold valuable information for ID classification. Hence, pre-trained models that have been trained to differentiate between various sound classes can be employed. An ID-classification model $f_s$ at the server is designed to predict IDs from the embeddings sent from each client. Additionally, each client is supplied with an anomaly detection model $f_{a, k}$, which is used to compute anomaly scores from the embeddings derived from both the pre-trained model $f_c$ and the ID-classification model $f_s$.

Firstly, each client $k$ computes embeddings $\mathbf{E}_k$ of raw data $\mathbf{X}_k$ using the pre-trained model $f_c$ and sends these embeddings along with the IDs $\mathbf{S}_k$ to the server. The calculated embeddings are anticipated to be beneficial for general audio classification and are also expected to include information helpful for ID classification.

Subsequently, the server receives the embeddings and IDs from all clients. It then trains the ID-classification model $f_s$ to predict the IDs from the embeddings. Contrary to the conventional outlier exposure approach where a classifier is trained to predict IDs based on raw data input, this ID-classification model predicts IDs based on the received embeddings. Once trained, this ID-classification model is distributed to all clients. 

Finally, at each client $k$, the embeddings $\mathbf{E}_k$ calculated from the raw data $\mathbf{X}_k$ are input into the received ID-classification model $f_s$ to generate new embeddings $\mathbf{\Tilde{E}}_k$. These new embeddings can be directly used for distinguishing audio data with different IDs, enabling the use of the outlier exposure approach.  If the embedding of an audio sample with a specific ID is closer to the embeddings of normal samples of that same ID rather than to the embeddings of normal samples from other IDs, the audio sample can be identified as normal. If not, it can be considered anomalous. Therefore, it is possible to apply anomaly detection methods such as k-NN (k-Nearest Neighbor) or LOF (Local Outlier Factor) to the computed embeddings to calculate anomaly scores and perform anomalous sound detection.  

\subsection{Difference from federated \& split learning}
Using the proposed method, even in extreme non-IID cases where each client has only a single machine operating under one condition, and each client's machine or operating condition is not identical, it is possible to collaboratively learn an ASD model using outlier exposure.
In contrast, in federated learning, the accuracy of the integrated model on the server side greatly decreases in extreme non-IID cases because the models learned at each client differ significantly. Also, in split learning, the accuracy of the model decreases significantly in extreme non-IID cases because the direction of updates of the model on the server side changes greatly for each client. In the proposed method, if the distributed pre-trained model can extract embeddings useful for the task that all clients collaborate to achieve, we can learn a high-accuracy model even in extreme non-IID cases.

\begin{algorithm}[t]  
    \SetAlgoLined  
    \DontPrintSemicolon  
    \caption{Collaborative ASD by embedding sharing with server side inference}\label{alg:fed_anomaly_detection_server_inference}  
    \SetAlgoVlined  
    \SetKw{In}{in}  
    \SetKw{Continue}{continue}  
    \KwIn{Same as in Algorithm 1} 
    \KwOut{Same as in Algorithm 1}  
    \BlankLine  
    \tcc{Training Phase}  
    \tcc{Clients calculate embeddings $\mathbf{E}_{k}$} 
    \tcc{Server trains $f_{s}$}  
    Same as in Algorithm 1
    \BlankLine
    \tcc{Server computes new embeddings}
    \For{each client $k$ }{  
        $\mathbf{\Tilde{E}}_{k} \leftarrow f_s (\mathbf{E}_{k})$ \\
        Train an anomaly detection model $f_{a,k}$ using embeddings $\mathbf{\Tilde{E}}_{k}$\\
    }  
    \tcc{Inference Phase}  
    \tcc{Client side} 
    \BlankLine  
    \For{each client $c$ in parallel}{
        $\mathbf{E}'_{k} \leftarrow f_c (\mathbf{X}'_{k})$ \\
        
        Send embeddings $\mathbf{E'}_{k}$ and IDs $\mathbf{S'}_k$ to the server\\ 
    }
    \tcc{Server side}  
    Server receives embeddings $\mathbf{E'}_{k}$ and IDs $\mathcal{S'}_k$ from each client $k$\\
    \For{each client $c$ }{  
        $\mathbf{\Tilde{E}'}_{k} \leftarrow f_s (\mathbf{E}'_{k})$ \\
        $\mathbf{A}_{k} \leftarrow f_a (\mathbf{\Tilde{E}'}_{k})$ \\
    }  
\end{algorithm}  
\subsection{Considerations for possible attacks}
We consider possible attacks against our proposed framework. 
Three types of attacks can be considered: sensitive information retrieval, raw data retrieval and model stealing. 

A sensitive information retrieval refers to the extraction of sensitive information retained within the shared embeddings.  If the server has access to both the shared embeddings and the pre-trained model $f_c$, it may be able to extract sensitive information from these shared embeddings. This can be achieved by preparing machine sound data tagged with sensitive information labels, calculating embeddings from the prepared data, and training a model to predict the sensitive information labels based on these embeddings.
However, in the proposed method, the server does not need access to the pre-trained model since it is only used on the clients. Therefore, sensitive information retrieval can be evaded by concealing the pre-trained model from the server. This can be accomplished either by agreeing on a pre-trained model between clients or by preparing a third-party distributor responsible for designing the pre-trained model.

Raw data retrieval refers to reconstructing raw data from the shared embeddings using a decoder model. The decoder model can be trained with access to sets of raw data and corresponding embeddings generated by the pre-trained model \cite{li2021learning}. However, it is possible to prevent raw data retrieval by concealing the pre-trained model from the server. Also, despite the fact that the reconstruction of raw data from embeddings has been studied in the context of text or images \cite{song2020information, li2021learning}, recreating noisy industrial machine sound data from embeddings can be particularly challenging. This is because raw data are typically contaminated by factory noise and the machine sound can drastically change based on the surrounding environments.

Model stealing refers to the unauthorized acquisition of trained models from either clients or the server.  
As the trained model can be considered an asset owned by the clients, its leakage can result in financial loss for them. In Algorithm \ref{alg:fed_anomaly_detection}, each client has access to both the pre-trained model and the ID-classification model. This implies that, if one of the clients is compromised, all models necessary for performing ASD could be leaked.
To mitigate this, Algorithm \ref{alg:fed_anomaly_detection_server_inference} could be employed. In this algorithm, the server trains the anomaly detection model and calculates anomaly scores. As a result, the ID-classification model will not be disclosed to clients, and compromising one client does not directly lead to the exposure of all models required for performing ASD. Additionally, if the pre-trained model is concealed from the server, compromising the server does not result in the exposure of all models required for ASD. However, as Algorithm \ref{alg:fed_anomaly_detection_server_inference} requires communication between the server and clients for calculating anomaly scores, the choice between Algorithm \ref{alg:fed_anomaly_detection} and Algorithm \ref{alg:fed_anomaly_detection_server_inference} should be determined based on practical considerations.

\section{Experiments}
\label{sec:typestyle}

\subsection{Dataset}
The DCASE 2020 Challenge Task2 development dataset \cite{Koizumi2020} was used for the experiments. Each audio file is a 10-second single-channel recording sampled at 16 kHz, originating from one of six types of machines (ToyCar, ToyConveyor, fan, pump, slider, valve). Each piece of data has a machine ID to distinguish between three to four machines of each type. This dataset was selected for its suitability in reproducing the specific scenario targeted in this study, as each machine ID in the dataset directly correspond to a unique machine.

\subsection{Experimental conditions}
We carried out experiments for each machine type, considering the machine ID as the corresponding ID for each sound data. In other words, we presumed an extreme non-IID situation where the number of clients equates to the number of machine IDs for each machine type. Each client possesses data from only one machine ID, and no two clients share data from the same machine ID.

We employed three pre-trained models as follows:

\noindent
\textbf{YAMNet.}
YAMNet \cite{google2020yamnet} is a pre-trained model trained on the audio event classification task using the AudioSet \cite{gemmeke2017audio}. For each 10-second audio file, we obtained embeddings comprised of 1024 dimensions and 20 frames. For the input into the MobileFaceNet, we stacked 8 frames, with an overlap of seven frames.

\noindent
\textbf{PANNs.}
PANNs \cite{kong2020panns} are pre-trained models also trained on the audio event classification task using the AudioSet. The specific architecture we employed from PANNs is the Wavegram-Logmel-CNN. In order to generate model inputs, we oversampled our 16 kHz data to create 32kHz data, which was then converted into mel spectrograms with 64 mel bins. For each audio file, we obtained embeddings of 2048 dimensions and 90 frames. For the input into the MobileFaceNet, we stacked 32 frames, with an overlap of 28 frames.

\noindent
\textbf{OpenL3.}
OpenL3 \cite{cramer2019look, arandjelovic2017look} is another pre-trained model trained by a self-supervised task that predicts whether the images and sounds obtained from AudioSet come from overlapping sections of the same video. We oversampled to 48kHz and calculated mel spectrograms with 256 mel bins. For each audio file, we created embeddings composed of either 512 or 6144 dimensions along with 96 frames. For the input into the MobileFaceNet, we stacked 32 frames with an overlap of 28 frames. We prepared two OpenL3 models: one with an embedding dimension of 512 (OpenL3-512) and the other with an embedding dimension of 6144 (OpenL3-6144).

We utilized MobileFaceNet \cite{chen2018mobilefacenets} as the ID-classification model, following \cite{morita2021anomalous}. We also incorporated arcmargin softmax loss \cite{deng2019arcface} in our model.
We trained the ID-classification model for 20 epochs using the AdamW optimizer \cite{loshchilov2019decoupled}, with a learning rate of 0.0001 and a batch size of 32.

For each embedding derived from the trained MobileFaceNet, we calculated the anomaly score using the k-NN method, with the number of neighbors set to 2.

We carried out experiments for each machine type and computed the ASD performance metrics for each machine ID. The area under the receiver operating characteristic curve (AUC) was used for the evaluation metric.

\begin{table*}[t]  
\begin{center}  
\caption{Anomaly detection results for different machine types. ``total'' denotes the average AUC for all 23 machine IDs. (OpenL3-512) and (OpenL3-6144) denote OpenL3 with embedding dimension of 512 and 6144, respectively.}  
\setlength\tabcolsep{8pt}
\begin{tabular}{l|c|c|c|c|c|c|c}  
\bhline{1.5pt}  
Model       &    {VAE}   &      {VIDNN} & {Glow} & {Proposed}& {Proposed} & {Proposed} & {Proposed} \\   
            &            &              &        & (YAMNet)  & (PANNs)        &(OpenL3-512) & (OpenL3-6144) \\ \hline    
ToyCar      &  83.3      &  81.2        & 75.9   & 78.9       & 79.0       & 87.5   & \bf{91.4}    \\   
ToyConveyor &  74.1      &  \bf{74.7}   & 70.7   & 63.4       & 63.1       & 67.8   & 70.2    \\   
fan         &  72.1      &  72.8        & 71.5   & 68.6      & 67.3       & 70.5   & \bf{76.4}    \\   
pump        &  76.2      &  76.6        & 76.3   & 71.1      & 76.0       & 81.6   & \bf{82.4}    \\   
slider      &  84.2      &  84.7        & 93.9   & 82.2      & 78.8       & \bf{96.2}   & 95.6    \\   
valve       &  70.2      &  83.0        & 86.1   & 89.4      & 89.6       & 94.3   & \bf{97.3}    \\ \bhline{1.5pt}   
total       &  76.8      &  79.0        & 79.4   & 76.1      & 76.2       & 83.7   & \bf{86.2}    \\ \bhline{1.5pt}    
\end{tabular}\\  
\label{result_1}  
\end{center}  
\end{table*}  
\subsection{Results}
Before calculating the ASD performance, we examined whether machine IDs can be predicted from the embeddings obtained by the pre-trained models. We used the trained MobileFaceNet to predict the machine ID of each normal sound file, and calculated the accuracy (ACC) of the prediction. We calculated ACC for each machine ID, and averaged all ACCs. Average ACCs were 77.83\% for YAMNet, 80.76\% for PANNs, 77.53\% for OpenL3-512, and 96.95\% for OpenL3-6144.
These results indicate that it is possible to predict machine IDs from the embeddings obtained by the pretrained models, and that the embeddings retain information necessary for ID-classification. However, the extent to which information necessary for ID-classification remains in the embeddings vary depending on the pre-trained models or the dimension of the embeddings. 

Table 1 presents the AUC for each machine type, computed by averaging the AUCs of each machine ID. The row labeled as "total" indicates the average AUC across all machine IDs. The results utilizing VAE, VIDNN, and Glow (normalizing flows) \cite{Kingma2018} were sourced from \cite{Dohi2021}. VAE, VIDNN \cite{Suefusa2020}, and Glow \cite{Kingma2018} are all unsupervised ASD models that were trained exclusively on local data from each respective client.

OpenL3-6144 attained the highest average AUC. It outperformed Glow, which was the best-performing method among those trained solely on each client's data, by 6.8\%. OpenL3-512 also achieved higher AUC than Glow, but its AUC was lower than OpenL3-6144. The proposed method using YAMNet and PANNs showed lower AUCs than other methods. This indicates that these models could have omitted crucial information required for outlier exposure. Regarding ToyConveyor, the AUCs of the proposed methods were inferior to Glow's. This observation is consistent with findings from other outlier exposure approaches \cite{Dohi2021}. In ToyConveyor, it is plausible that the boundaries identifying machine ID and those distinguishing between normal and anomaly data may differ.

\section{Conclusion}
We proposed an ASD method that enables multiple clients to collaborate on learning an ASD model while keeping each other's raw data confidential. The proposed method works even in non-IID cases where the distribution of each client's data differs greatly. In the proposed method, we calculated embeddings at each client using a pre-trained model to identify sound data, and aggregated the calculated embeddings on the server to perform ASD by outlier exposure. The proposed method using OpenL3 as the pre-trained model improved the AUC by an average of 6.8\%.

\bibliographystyle{IEEEbib}
\bibliography{refs}

\begin{thebibliography}{10}

\bibitem{Koizumi2017}
Y.~{Koizumi}, S.~{Saito}, H.~{Uematsu}, and N.~{Harada},
\newblock ``Optimizing acoustic feature extractor for anomalous sound detection based on neyman-pearson lemma,''
\newblock in {\em EUSIPCO}, 2017, pp. 698--702.

\bibitem{Suefusa2020}
K.~{Suefusa}, T.~{Nishida}, H.~{Purohit}, R.~{Tanabe}, T.~{Endo}, and Y.~{Kawaguchi},
\newblock ``Anomalous sound detection based on interpolation deep neural network,''
\newblock in {\em ICASSP}, 2020, pp. 271--275.

\bibitem{mnasri2022anomalous}
Z.~Mnasri, S.~Rovetta, and F.~Masulli,
\newblock ``Anomalous sound event detection: A survey of machine learning based methods and applications,''
\newblock {\em Multimedia Tools and Applications}, vol. 81, pp. 5537--5586, 2022.

\bibitem{Giri2020}
R.~Giri, S.~V. Tenneti, K.~Helwani, F.~Cheng, U.~Isik, and A.~Krishnaswamy,
\newblock ``Unsupervised anomalous sound detection using self-supervised classification and group masked autoencoder for density estimation,''
\newblock Tech. {R}ep., DCASE2020 Challenge, 2020.

\bibitem{zhao2018federated}
Yue Zhao, Meng Li, Liangzhen Lai, Naveen Suda, Damon Civin, and Vikas Chandra,
\newblock ``Federated learning with non-iid data,''
\newblock {\em arXiv preprint arXiv:1806.00582}, 2018.

\bibitem{mcmahan2016communication}
H.~Brendan McMahan, Eider Moore, Daniel Ramage, Seth Hampson, and Blaise Ag{\"u}era~y Arcas,
\newblock ``Communication-efficient learning of deep networks from decentralized data,''
\newblock {\em arXiv preprint arXiv:1602.05629}, 2016.

\bibitem{vepakomma2018split}
Praneeth Vepakomma, Otkrist Gupta, Tristan Swedish, and Ramesh Raskar,
\newblock ``Split learning for health: Distributed deep learning without sharing raw patient data,''
\newblock in {\em NeurIPS}, 2018.

\bibitem{Kingma2014a}
D.~P. Kingma and Max Welling,
\newblock ``Auto-encoding variational bayes,''
\newblock in {\em ICLR}, 2014.

\bibitem{Tabak2013}
E.~Tabak and C.~Turner,
\newblock ``A family of nonparametric density estimation algorithms,''
\newblock {\em Communications on Pure and Applied Mathematics}, vol. 66, pp. 145--164, 2013.

\bibitem{Kingma2018}
D.~P. Kingma and P.~Dhariwal,
\newblock ``Glow: Generative flow with invertible 1x1 convolutions,''
\newblock in {\em NeurIPS}, 2018.

\bibitem{Koizumi2020spider}
Yuma Koizumi, Masahiro Yasuda, Shin Murata, Shoichiro Saito, Hisashi Uematsu, and Noboru Harada,
\newblock ``Spidernet: Attention network for one-shot anomaly detection in sounds,''
\newblock in {\em ICASSP}, 2020, pp. 281--285.

\bibitem{gao2020endtoend}
Yansong Gao, Minki Kim, Sharif Abuadbba, Yeonjae Kim, Chandra Thapa, Kyuyeon Kim, Seyit~A. Camtepe, Hyoungshick Kim, and Surya Nepal,
\newblock ``End-to-end evaluation of federated learning and split learning for internet of things,''
\newblock in {\em Proceedings of the 39th International Symposium on Reliable Distributed Systems (SRDS)}, 2020.

\bibitem{li2021learning}
Chu-Chen Li, Cheng-Te Li, and Shou-De Lin,
\newblock ``Learning privacy-preserving embeddings for image data to be published,''
\newblock {\em ACM Transactions on Intelligent Systems and Technology}, vol. 14, no. 6, pp. 1--26, 2021.

\bibitem{song2020information}
Congzheng Song and Ananth Raghunathan,
\newblock ``Information leakage in embedding models,''
\newblock in {\em Proceedings of the 2020 ACM SIGSAC Conference on Computer and Communications Security}. 2020, ACM.

\bibitem{Koizumi2020}
Y.~Koizumi, Y.~Kawaguchi, K.~Imoto, T.~Nakamura, Y.~Nikaido, R.~Tanabe, H.~Purohit, K.~Suefusa, T.~Endo, M.~Yasuda, and N.~Harada,
\newblock ``Description and discussion on {DCASE}2020 {C}hallenge task2: Unsupervised anomalous sound detection for machine condition monitoring,''
\newblock in {\em DCASE Workshop}, 2020.

\bibitem{google2020yamnet}
Google~AI team,
\newblock ``Yamnet,'' Available at: \url{https://github.com/tensorflow/models/tree/master/research/audioset/yamnet}, 2020.

\bibitem{gemmeke2017audio}
Jort~F Gemmeke, Daniel~PW Ellis, Dylan Freedman, Aren Jansen, Wade Lawrence, R~Channing Moore, Manoj Plakal, and Marvin Ritter,
\newblock ``Audio set: An ontology and human-labeled dataset for audio events,''
\newblock in {\em ICASSP}. IEEE, 2017, pp. 776--780.

\bibitem{kong2020panns}
Qiuqiang Kong, Yin Cao, Turab Iqbal, Yuxuan Wang, Wenwu Wang, and Mark~D. Plumbley,
\newblock ``Panns: Large-scale pretrained audio neural networks for audio pattern recognition,''
\newblock {\em IEEE/ACM Transactions on Audio, Speech, and Language Processing}, vol. 28, 2020.

\bibitem{cramer2019look}
Aurora Cramer, Ho-Hsiang Wu, Justin Salamon, and Juan~Pablo Bello,
\newblock ``Look, listen and learn more: Design choices for deep audio embeddings,''
\newblock in {\em ICASSP}, 2019, pp. 3852--3856.

\bibitem{arandjelovic2017look}
Relja Arandjelović and Andrew Zisserman,
\newblock ``Look, listen and learn,''
\newblock in {\em IEEE International Conference on Computer Vision (ICCV)}, 2017.

\bibitem{chen2018mobilefacenets}
Sheng Chen, Yang Liu, Xiang Gao, and Zhen Han,
\newblock ``Mobilefacenets: Efficient cnns for accurate real-time face verification on mobile devices,''
\newblock {\em arXiv preprint arXiv:1804.07573}, 2018.

\bibitem{morita2021anomalous}
Kazuki Morita, Tomohiko Yano, and Khai~Q Tran,
\newblock ``Anomalous sound detection using cnn-based features by self supervised learning,''
\newblock Tech. {R}ep., DCASE2021 Challenge, 2021.

\bibitem{deng2019arcface}
Jiankang Deng, Jia Guo, Niannan Xue, and Stefanos Zafeiriou,
\newblock ``Arcface: Additive angular margin loss for deep face recognition,''
\newblock in {\em Proceedings of the IEEE/CVF Conference on Computer Vision and Pattern Recognition}, 2019, pp. 4690--4699.

\bibitem{loshchilov2019decoupled}
Ilya Loshchilov and Frank Hutter,
\newblock ``Decoupled weight decay regularization,''
\newblock in {\em ICLR}, 2019.

\bibitem{Dohi2021}
K.~Dohi, T.~Endo, H.~Purohit, R.~Tanabe, and Y.~Kawaguchi,
\newblock ``Flow-based self-supervised density estimation for anomalous sound detection,''
\newblock in {\em ICASSP}, 2021.

\end{thebibliography}
\end{document}